\documentclass[aip,pop,preprint]{revtex4-1}
\usepackage{amsmath}
\usepackage{amssymb}
\usepackage{color}
\usepackage{epsfig}
\usepackage{graphicx}
\usepackage{bm}
\usepackage{natbib}
\usepackage{algorithmic}
\usepackage{algorithm}
\usepackage{tabularx}
\usepackage{url}
\usepackage[utf8]{inputenc}
\usepackage[T1]{fontenc}
\usepackage{mathptmx}

\begin{document}

\preprint{AIP/123-QED}

\title{The Reversibility Of Magnetic Reconnection}

\author{M. Xuan}
\author{M. Swisdak}
\email{swisdak@umd.edu}
\author{J. F. Drake}
\affiliation{Institute for Research in Electronics and Applied Physics,
University of Maryland, College Park, MD 20742}

\date{\today}

\begin{abstract}

The reversibility of the transfer of energy from the magnetic field to
the surrounding plasma during magnetic reconnection is examined.
Trajectories of test particles in an analytic model of the fields
demonstrate that irreversibility is associated with separatrix
crossings and regions of weaker magnetic field.  Inclusion of a guide
field increases the degree of reversibility.  Full kinetic simulations
with a particle-in-cell code support these results and demonstrate
that while time-reversed simulations at first ``un-reconnect'', they
eventually evolve into a reconnecting state.

\end{abstract}

%\pacs{52.65.-y,52.65.Rr,52.27.Ny}

\maketitle

\section{Introduction}\label{intro}

The mean free path associated with classical collisions significantly
exceeds the typical scale sizes in many heliospheric and astrophysical
plasmas.  This simple observation raises interesting questions as to
how energy dissipates in these systems, particularly for phenomena
such as magnetic reconnection \cite{drake06a, hesse11a, cassak16a},
turbulence \cite{cranmer03a}, and shocks \cite{krall97a} in which
kinetic scales are important.  In the complete absence of collisions,
Boltzmann's equation describing the evolution of the distribution
function $f$ reduces to the Vlasov equation, which is formally
reversible in time.  Reversibility implies, among other things, that
in a closed system the entropy -- and in fact any function of $f$ --
remains constant during the system's evolution.  Despite this
restrictive mathematical constraint, Landau \cite{landau46a}
demonstrated that one-dimensional electrostatic plasma waves in the
nominally collisionless limit are accompanied by a cascade to smaller
and smaller structures in phase space which, in an actual plasma,
inevitably lead to dissipation, irreversibility, and an increase in
entropy.  Landau damping has been experimentally confirmed
\cite{malmberg64a} and subsequent work has shown that the same basic
mechanism occurs in association with many other plasma oscillations
\cite{wong64a,ryutov99a}.

The example of Landau damping demonstrates that mathematically
reversible processes can, in fact, lead to irreversibility in real
systems.  An interesting question is whether that conclusion extends
to the phenomenon of magnetic reconnection.  During reconnection a
change in magnetic topology triggers a transfer of energy from the
field to the surrounding plasma as slowly inflowing plasma crosses
magnetic separatrices and is accelerated to Alfv\'enic velocities
flowing away from an X-point \cite{zweibel09a}.  In a strictly
collisionless Vlasov-governed system the transfer of energy from field
to particles is completely reversible, implying that it should be
possible for reconnection to run backwards, with Alfv\'enic plasma
jets converging at the X-point accompanied by a transfer of energy
from the plasma to the magnetic field (i.e., reversed reconnection
would act as a dynamo).  This sequence of events is not observed in
nature, suggesting that, in practice if not in theory, collisionless
magnetic reconnection is irreversible.

Data from the Magnetospheric Multiscale (MMS) Mission have
demonstrated close agreement between in situ observations of
reconnection and the results of particle-in-cell (PIC) simulations
\cite{burch16a,burch17a,price19a}.  Such simulations follow
macroparticles as they move through their collective self-consistent
electromagnetic fields interpolated to a numerical grid and, in
principle, capture all of the important dynamics in reconnecting
plasmas.  Since, as with the (assumed to be collisionless) physical
system, the equations evolved in PIC codes are time-reversible,
examining questions concerning reversibility in PIC simulations serves
as a useful proxy for actual reconnection.  We will show below that
reconnection behaves irreversibly in such simulations.  In Section
\ref{test} we discuss test-particle trajectories in a simple analytic
model of reconnection.  Section \ref{pic} describes self-consistent
particle-in-cell simulations of reconnection run forward and backward
in time while Section \ref{disc} discusses the results and their
broader implications.

\section{Test Particle Trajectories}\label{test}

Particle trajectories in self-consistent PIC simulations will be
discussed below, but we first consider a simpler model that captures
the key features of fields near a reconnection X-point.  Observations
and simulations have demonstrated that reconnection in a fully
three-dimensional domain is often accompanied by turbulence and
complicated field line trajectories \cite{daughton11a,burch17a}.
Despite the complexity, however, many important features of
reconnection, including the large-scale topology and the reconnection
rate, remain similar to what is observed in simulations with reduced
domains, sometimes called 2.5D, in which variations perpendicular to
the reconnection plane are suppressed \cite{hesse01a}.  Since, as we
will show, reconnection is irreversible in such restricted domains it
seems clear that it will also be irreversible in a fully
three-dimensional system.

\subsection{Model}\label{model}

With that motivation we seek a simple model of a reconnection X-point
in a system with an invariant direction, here parallel to
$\mathbf{\hat{y}}$.  Two cases are of interest: Anti-parallel
reconnection, in which the reconnecting fields, which lie parallel to
$\mathbf{\hat{x}}$, have a shear angle of $180^{\circ}$; and
guide-field reconnection in which a spatially constant out-of-plane
component of the magnetic field reduces the shear angle.

We consider a dimensionless model, i.e., one in which every variable
is associated with a suppressed scaling factor that carries
information about the proper units.  Let the only non-zero component
of the vector potential have the form
\begin{equation}\label{vecpot}
A_y = -\frac{1}{2}\left(z^2-\alpha_B x^2\right) -\alpha_E t
\end{equation}
where $\alpha_B$ and $\alpha_E$ are positive constants and the
electrostatic potential $\phi=0$.  The magnetic and electric fields
are then
\begin{equation}
\mathbf{B} = z\mathbf{\hat{x}} + \alpha_B x\mathbf{\hat{z}} \qquad
\mathbf{E} = \alpha_E\mathbf{\hat{y}}
\end{equation}
Since $E_y$, which is equivalent to the reconnection rate, is constant
in space, Faraday's Law and the invariance with respect to $y$ imply
$\mathbf{B}$ is stationary in time.  No guide field is present so the
unreconnected fields on either side of the current layer are
anti-parallel and $E_{\parallel} =
(\mathbf{E}\boldsymbol{\cdot}\mathbf{B})/B$ trivially vanishes
everywhere.  Ellipses centered at the origin mark contours of constant
magnetic field strength.  Magnetic separatrices along the lines $z =
\pm \sqrt{\alpha_B}x$ divide space into upstream ($|z| >
\sqrt{\alpha_B}\,|x|$) and downstream ($|z| < \sqrt{\alpha_B}\,|x|$)
regions.  Similar configurations have previously been investigated as
models of reconnection \cite{craig94a,nickeler19a}.

Generally speaking, the two instances of $\alpha$ in equation
\ref{vecpot} can take different values with $\alpha_{\text{B}}$
related to the angle of the separatrices and $\alpha_{\text{E}}$ to
the reconnection rate.  (If, for example, $\alpha_{\text{B}}=1$ and
$\alpha_E=0$ then the field lines meet at right angles at the X-point,
the associated current vanishes, and reconnection does not occur.)
However, \cite{liu17a} has argued that constraints at
magnetohydrodynamic scales impose the condition $\alpha_{\text{E}}
\approx \alpha_{\text{B}}$ when $\alpha_{\text{B}}\ll 1$.  In this
work we choose $\alpha_{\text{B}} = \alpha_{\text{E}} = \alpha = 0.1$.

\begin{figure}
\includegraphics[width=0.7\textwidth]{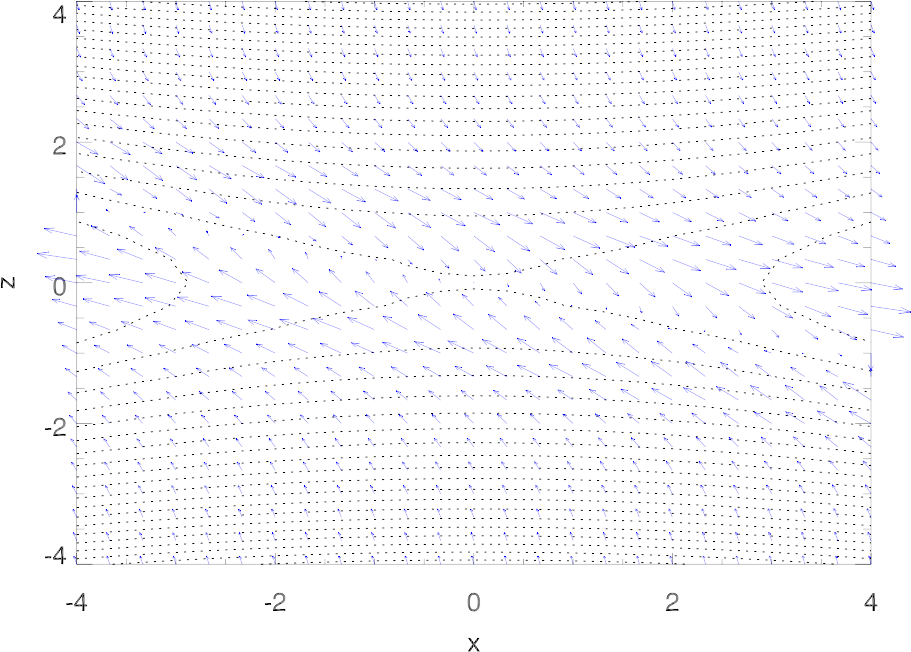}
\caption{\label{psi_exb} Magnetic field lines (dashed) and in-plane
  $\mathbf{E}\boldsymbol{\times}\mathbf{B}$ flow vectors (solid blue)
  for the analytic model with $\alpha = 0.1$ and $B_g = 1$.}
\end{figure}

The symmetry inherent in the reconnection of anti-parallel fields is
broken by an out-of-plane (guide) component, which can in practice have
significant effects even when relatively small in magnitude
\cite{swisdak05a}.  Including a constant $B_y \neq 0$ while
maintaining the condition $E_{\parallel} =0$ implies the existence of
non-zero in-plane electric fields.  An electrostatic potential $\phi$,
with $\mathbf{E} = -\boldsymbol{\nabla}\phi$, generating these
components satisfies
\begin{equation}
-z\frac{\partial \phi}{\partial x} - \alpha x\frac{\partial
  \phi}{\partial z} + \alpha B_y = 0
\end{equation}
the solution to which can be found from the method of characteristics to be 
\begin{equation}\label{phi}
\phi = \sqrt{\alpha} B_y\log|z+\sqrt{\alpha}x| + g(z^2 - \alpha x^2)
\end{equation}
with $g$ an arbitrary function that can, in general, be chosen to
satisfy boundary conditions but will be neglected here.  The resulting
in-plane components of $\mathbf{E}$ are
\begin{equation}\label{inplanee}
E_x = -\alpha\frac{B_y}{z+\sqrt{\alpha}x} \qquad
E_z = -\sqrt{\alpha}\frac{B_y}{z+\sqrt{\alpha}x}
\end{equation}
In-plane contours of the $\mathbf{E}\boldsymbol{\times}\mathbf{B}$
flow and magnetic field lines for the case $B_y=1$ are shown in Figure
\ref{psi_exb}.  Interestingly, despite the simplicity of the model,
there are striking similarities to particle-in-cell simulations of
guide-field reconnection \cite{pritchett01b}, including strong flows
across the $z=\sqrt{\alpha}x$ separatrix and weak sheared flows
straddling $z=-\sqrt{\alpha}x$.

The existence of singularities in $\phi$, $E_x$, and $E_z$ along the
$z = -\sqrt{\alpha}x$ separatrix has deeper roots than the specific
choices of functional forms assumed in this model.  First note that it
is not possible to choose a $g$ in equation \ref{phi} that eliminates
the singularities.  Since $g$ is only a function of $z^2-\alpha x^2$,
it has the same magnitude on both $z=\sqrt{\alpha}x$ and
$z=-\sqrt{\alpha}x$ so that any cancellation along the latter will
create a new singularity on the former.  More generally, any 2.5D
model with a time-stationary magnetic field that includes a spatially
constant out-of-plane component and satisfies
$\mathbf{E}\boldsymbol{\cdot}\mathbf{B}=0$ must have singularities
associated with the vanishing of the in-plane field.  In such a model
$B_x=B_z=0$ at the X-point, making it impossible to satisfy the other
requirements there while simultaneously maintaining $E_{\parallel}=0$.
When the model includes a guide field in this work we only consider
particle trajectories that do not cross the $z=-\sqrt{\alpha}x$
separatrix.

\subsection{Numerical Results}\label{trajs}

In order to explore the reversibility of individual trajectories, we
introduce test particles into the model's magnetic and electric
fields.  The field configuration remains fixed as the positions and
velocities of the particles are integrated forward in time using the
non-relativistic Newton-Lorentz equations: $d\mathbf{x}/dt =
\mathbf{v}$ and $d\mathbf{v}/dt = q(\mathbf{E} +
\mathbf{v}\boldsymbol{\times}\mathbf{B})$.  These equations are
reversible and so particles followed forward and backward in time
should return to their initial locations.  (As in Section \ref{model},
the variables are assumed to include a suppressed factor representing
the dimensional information.  For instance, $q=\pm 1$ for protons or
electrons respectively, although in what follows we choose $q=1$ for
simplicity.)

In practice, however, the accumulation of small integration errors has
the potential to alter trajectories.  These alterations will be of
particular significance when particles pass near locations such as
X-points or separatrices where the $\kappa^2$ parameter, introduced in
\cite{beuchner89a} and given by the ratio of the radius of curvature
of the magnetic field to the particle Larmor radius, most closely
approaches values associated with chaotic motion\cite{escande00a}.
Because X-points and separatrices do not fill the domain, the
trajectory of every particle that passes near one effectively threads
a needle such that any small change can lead to significant deviations
when the orbit is tracked backwards in time.  Here the deviations
arise from numerical errors in solving the governing equations, but in
real plasmas analogous behavior can arise from, for instance,
small-amplitude electric field fluctuations.  To emphasize the analogy
with Landau damping: During reconnection in either an actual plasma or
a numerical simulation there will always be random fluctuations that
sufficiently perturb particle trajectories and destroy fine-scale
phase-space structures.  When such fluctuations coincide with regions
associated with chaotic particle motions, irreversible particle
trajectories result.

We follow particle trajectories via a trapezoidal-leapfrog method with
the velocity integration employing the Boris algorithm
\cite{boris70a,birdsall91a,zeiler02a}.  The analogy with Landau
damping argues that exact numerical convergence is impossible
(achieving it would effectively imply infinite phase-space
resolution), in that for any timestep there will always exist
particles whose trajectories cannot be exactly reversed.  In this work
we have chosen a timestep, $\Delta t = 10^{-3}$, for which
trajectories that do not pass close to the separatrices and X-point
are reversible, but emphasize that tests with different values of
$\Delta t$ yield qualitatively similar results.  In general, reducing
$\Delta t$ will increase the number of particles and length of time
for which trajectories are reversible but will not eliminate
irreversibility altogether.  Here we show trajectories evolved forward
in time for $2.5 \times 10^5$ steps and then reversed for an equal
period.

\begin{figure}
\includegraphics[width=0.7\textwidth]{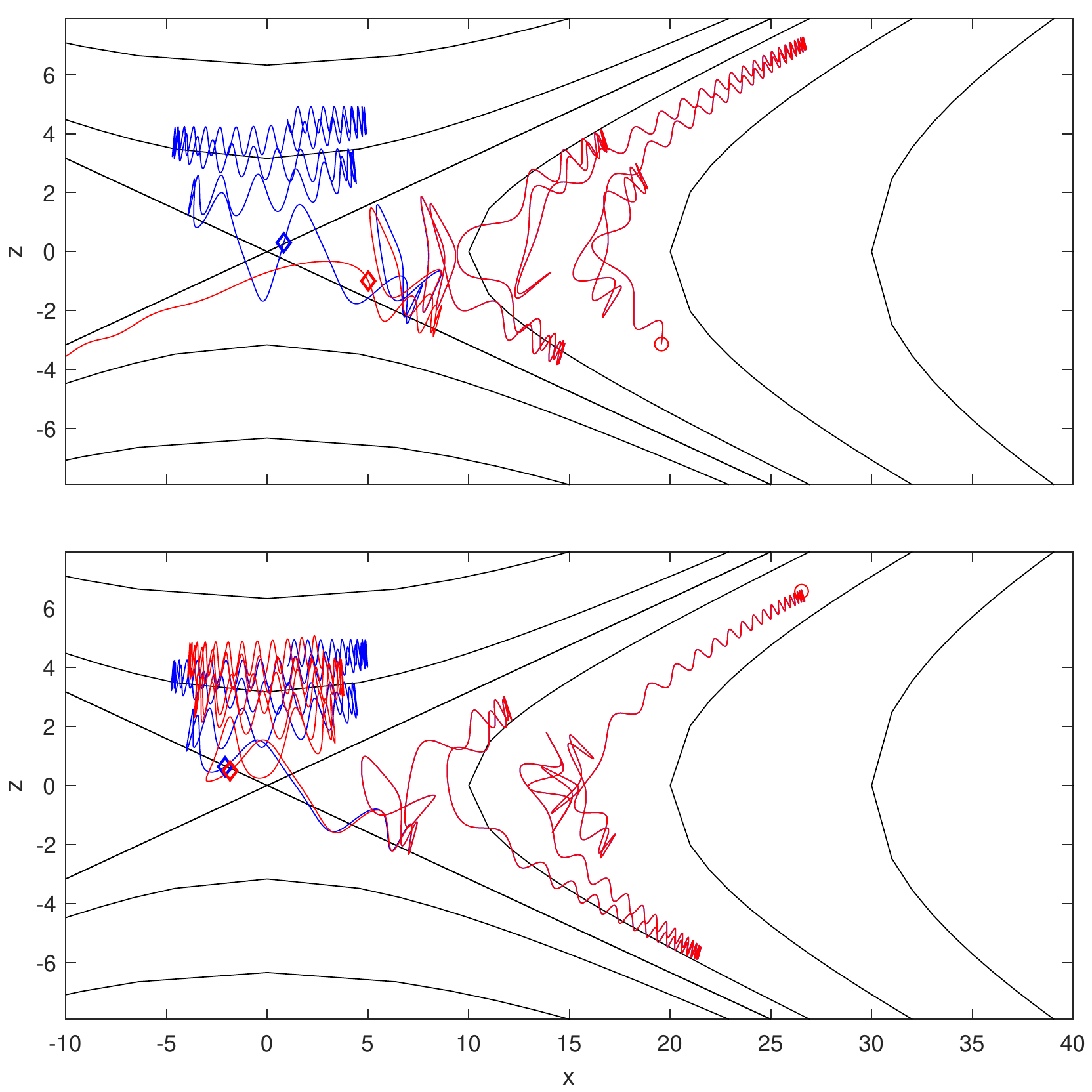}
\caption{\label{single_anti} Magnetic field lines (solid black) and
  two particle trajectories in the analytic model with $\alpha = 0.1$
  and $B_y = E_x=E_z=0$.  Blue lines represent the particle
  trajectories taken forward in time, red when reversed.  Each
  particle begins on a gyro-orbit centered at $x=1$, $z=4.5$ but
  separated in phase by $\pi/2$.  The locations where the reversals
  occur are marked by circles.  The diamonds indicate the locations
  discussed in Figure \ref{mu}.  [Associated dataset available at
    \protect\url{http://dx.doi.org/10.5281/zenodo.4608531}.]\cite{reversibility_fig2}}
\end{figure}

Figure \ref{single_anti} shows two particle trajectories in the
anti-parallel case of the model described in Section \ref{model}.  At
$t=0$ both have representative thermal velocities, with the
perpendicular velocity augmented with the local
$\mathbf{E}\boldsymbol{\times}\mathbf{B}$ drift.  The particles have
the same guiding centers but different phases (offset by $\pi/2$) so
that they initially occupy different locations on the same Larmor
orbit centered at $x=1$, $z=4.5$.  Both particles begin upstream and
are tracked forward in time (blue lines) as they pass near the X-line
and head downstream until, at $t=250$, the particles have the
locations given by the colored circles (top: $(x,z) \approx (20,-3)$;
bottom: $(x,z)\approx (27,6)$).  At the beginning of their
trajectories the particles exhibit the expected helical motions about
the field with a pitch given by $v_{\parallel}/v_{\perp}$.
Conservation of the adiabatically invariant magnetic moment implies
that $v_{\perp}$ varies in proportion to $\sqrt{B}$ so that the pitch
decreases (i.e., the helix winds more tightly) at the mirror points
and then increases as the particle moves back towards the origin.
Magnetic moment conservation begins to break down near the current
layer at $z=0$ due to the decrease in $B$.  Despite having initial
conditions on the same Larmor orbit, the blue trajectories clearly
differ near the X-point.  Once the particles move away from the
current layer they re-magnetize, mirror in the increasing field, and
again pass through the region of small field.  The overall behavior is
quite similar to that seen in particle trajectories taken from full
PIC simulations of reconnection \cite{chen08a,egedal19a}.

At $t=250$ the particle positions are marked by circles and their
trajectories are then integrated backwards in time (red lines) by
replacing $\Delta t$ with $-\Delta t$ in the discretization of the
Newton-Lorentz equations.  (Mathematically, identical results are
found by keeping $\Delta t$ unchanged but mapping $\mathbf{v}
\rightarrow -\mathbf{v}$ and either $\mathbf{B} \rightarrow
-\mathbf{B}$ or both $q\rightarrow -q$ and $\mathbf{E} \rightarrow
-\mathbf{E}$.)  While both particles at first closely re-trace their
inbound trajectories -- the blue lines at first cannot be seen because
they coincide with the red ones -- each eventually acquires a
significant deviation after which the incoming and outgoing
trajectories are more or less independent. 

\begin{figure}
\includegraphics[width=0.7\textwidth]{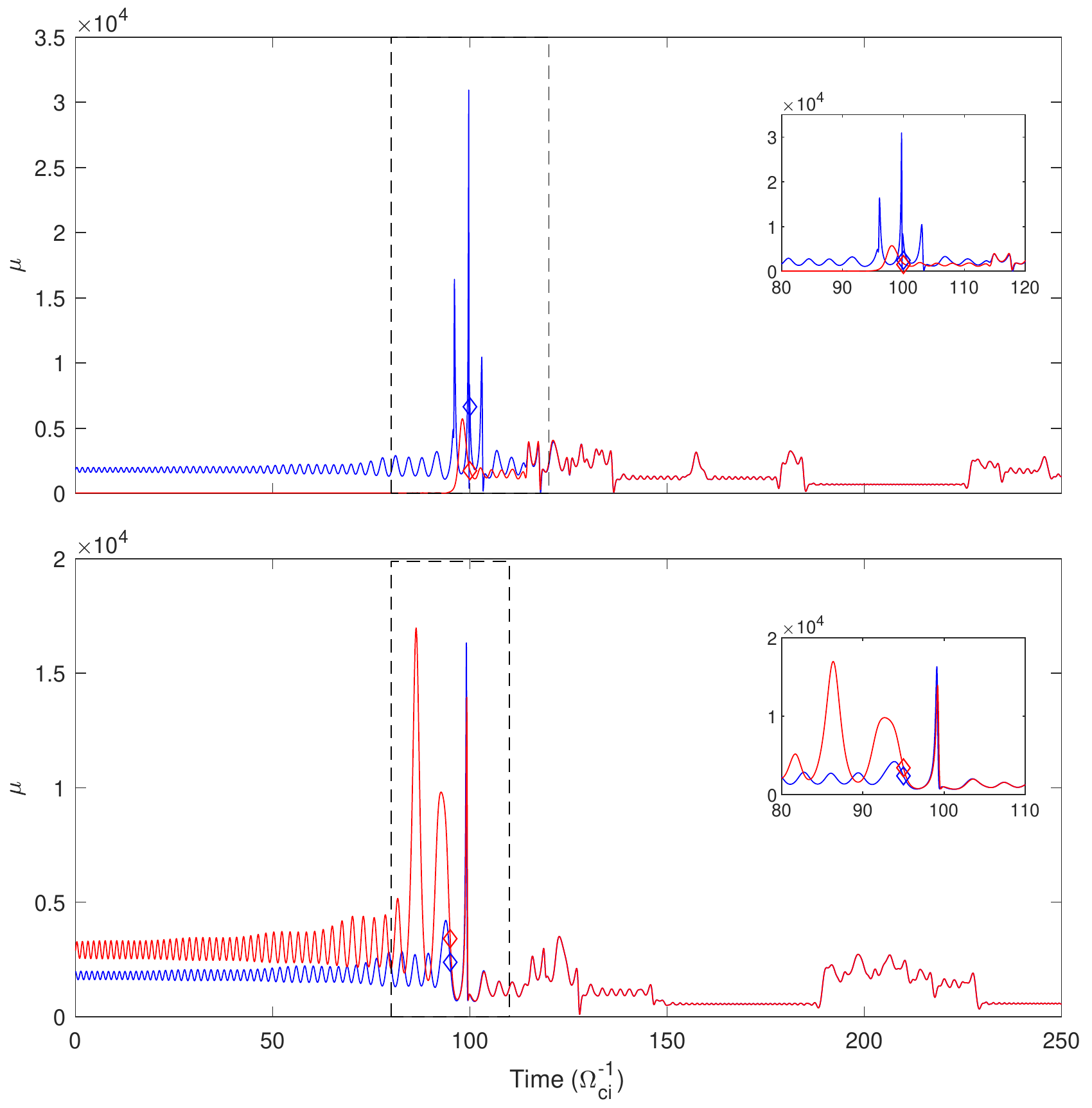}
\caption{\label{mu} The first adiabatic invariant $\mu$ as a function
  of time for the particle trajectories shown in Figure
  \ref{single_anti}.  The forward trajectory is shown in blue and the
  reverse in red.  The inset of each panel shows a blow-up of the
  critical region where the forward and backward phases begin to
  differ significantly.  The red and blue diamonds label the points
  shown marked similarly in Figure \ref{single_anti}.  [Associated
    dataset available at
    \protect\url{http://dx.doi.org/10.5281/zenodo.4608541}.]\cite{reversibility_fig3}}
\end{figure}

Figure \ref{mu} shows $\mu$, the first adiabatic invariant, for each
particle as a function of time.  As in Figure \ref{single_anti}, $\mu$
during the forward trajectory is shown in blue, while that of the
reverse is plotted in red.  The large spikes correspond to times when
the particles approach the X-point and $B$ approaches zero.  The
inset in each panel displays the times where the forward and backward
values notably begin to deviate and the diamonds correspond to those
plotted in Figure \ref{single_anti}.  In the first case the diamonds
are shown slightly after the deviation becomes noticeable at $t=110$
when the particles are oscillating across the current sheet at $(x,z)
\approx (7,-1)$.

\begin{figure}
\includegraphics[width=0.7\textwidth]{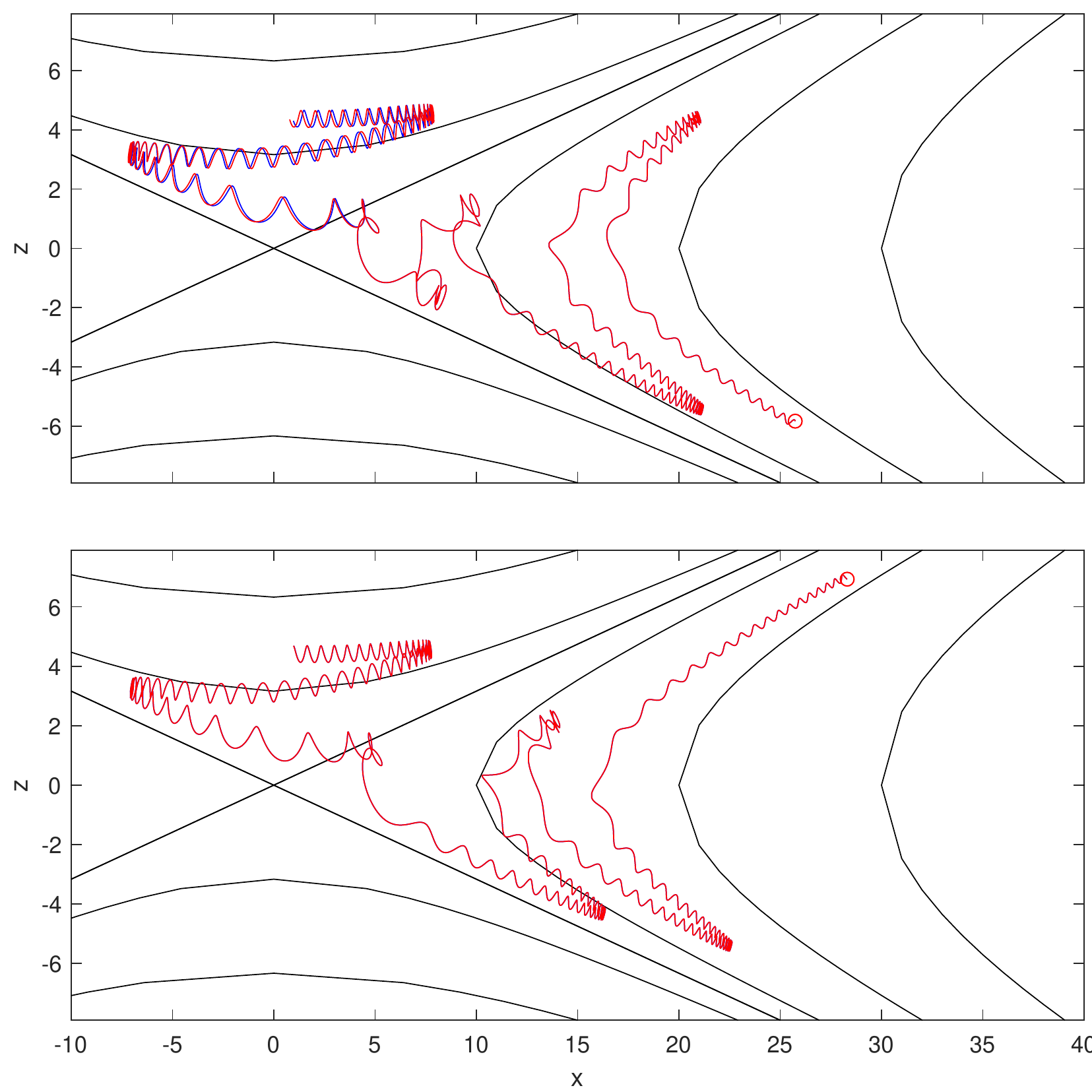}
\caption{\label{single_guide} Magnetic field lines (solid black) and
  two particle trajectories in the analytic model with $\alpha = 0.1$,
  $B_y = 1$, and $E_x$ and $E_z$ given by equation \ref{inplanee} in
  the same format as Figure \ref{single_anti}.  The particles begin on
  a gyro-orbit centered at $x=1$, $z=4.4$.  In the bottom panel the
  red reversed trajectory cover the blue forward trajectory.
  [Associated dataset available at
    \protect\url{http://dx.doi.org/10.5281/zenodo.4608551}.]\cite{reversibility_fig4}}
\end{figure}

Figure \ref{single_guide} shows a similar plot for a configuration
that includes a uniform guide field $B_y=1$ and the associated
in-plane electric fields given in equation \ref{inplanee}.  As in the
anti-parallel case, the two particles begin with the same velocities
and guiding centers (centered at $x=1$, $z=4.4$) but different
gyrophases.  The Larmor orbits again exhibit variations in pitch as
the particles encounter regions where $B$ changes.  Notably, however,
the presence of the guide field means that the particles' magnetic
moment is mostly conserved, the $\kappa^2$ parameter of
\cite{beuchner89a} remains in the non-chaotic regime, and the
particles never fully de-magnetize.  As a consequence, passing through
the central current sheet does not lead to significant differences in
the trajectories as it did in the anti-parallel case.  Furthermore, as
expected, the reversed-in-time trajectories more closely track the
forward-in-time motion.  In the bottom panel the reversal is nearly
exact, with the red reversed trajectory essentially covering the
forward blue trajectory.

The results shown in Figures \ref{single_anti} and \ref{single_guide}
suggest that the degree of irreversibility is higher in the
anti-parallel case.  It should be emphasized that the plotted
trajectories are meant to be suggestive but not representative.
Changes in multiple factors -- numerical method, length of
integration, timestep, initial position or velocity -- will produce
different trajectories.  A more comprehensive approach comes from
examining self-consistent simulations of reconnection.

%\begin{figure}
%\includegraphics[width=0.8\textwidth]{2both_26_crop}
%\caption{\label{single_anti} Magnetic field lines (solid) and two
%  particle trajectories in the analytic model with $\alpha = 0.1$ and
%  $B_y = E_x=E_z=0$.  Solid lines represent the particle trajectories
%  taken forward in time, dashed when reversed.  Both particles begin
%  on a gyro-orbit centered at $x=1$, $z=5$.  The locations where the
%  reversal occurs and the ends of the backwards integrations are marked
%  by circles and diamonds respectively.}
%\end{figure}

%\begin{figure}
%\includegraphics[width=0.7\textwidth]{phase_plot_marker}
%\caption{\label{phase} Accumulated phase as a function of time for the
%  particle trajectories shown in Figure \ref{single_anti}.  The inset
%  in the bottom right of each panel shows a blow-up of the critical
%  region where the forward and backward phases begin to differ
%  significantly.  The red and blue diamonds label the points shown
%  marked similarly in Figure \ref{single_anti}.}
%\end{figure}

%\begin{figure}
%\includegraphics[width=0.8\textwidth]{2both_26_guide_crop}
%\caption{\label{single_guide} Magnetic field lines (solid) and two
%  particle trajectories in the analytic model with $\alpha = 0.1$,
%  $B_y = 1$, and $E_x$ and $E_z$ given by equation \ref{inplanee} in
%  the same format as Figure \ref{single_anti}.  The particles begin on
%  a gyro-orbit centered at $x=-10.25$, $z=4$}
%\end{figure}

\section{Particle-in-Cell Simulations}\label{pic}

Particle-in-cell simulations follow the motions of many plasma
particles in the self-consistent fields they generate.  A key question
is whether the results found in Section \ref{test} for individual
trajectories continue to apply.  We perform such simulations with the
code {\tt p3d} \citep{zeiler02a}.  In its normalization a reference
magnetic field strength $B_0$ and density $n_0$ define the velocity
unit $v_{A0}=B_0/\sqrt{4\pi m_in_0}$ where $m_i$ is the ion mass.
Times are normalized to the inverse ion cyclotron frequency
$\Omega_{i0}^{-1}=m_ic/eB_0$, lengths to the ion inertial length
$d_{i0} =c/\omega_{pi0}$ (where $\omega_{pi0} = \sqrt{4\pi n_0
  e^2/m_i}$ is the ion plasma frequency), electric fields to
$v_{A0}B_0/c$, and temperatures to $m_iv_{A0}^2$.

The system parameters follow those of the GEM reconnection challenge
\cite{birn01a}.  The computational domain measures $L_x\times L_z =
25.6 \times 12.8 d_i$.  The ion-to-electron mass ratio is taken to be
$25$, which is sufficient to separate the electron and ion scales (the
electron inertial length $d_{e0} = 0.2d_{i0}$).  The electron thermal
speed $v_{th,e} \approx 2$ is much less than the normalized speed of
light, $c=20$; the latter further implies that
$\omega_{pe}/\Omega_{ce}=4$.  The spatial grid has resolution $\Delta
= 0.05$ in normalized units while the Debye length, $\approx 0.04$, is
the smallest physical scale.  The timestep is $\Delta t = 0.001$.
Each grid cell in the initial current sheet contains $\approx 400$
macroparticles.  The boundaries are periodic in the horizontal
direction and conducting perfectly reflecting walls at the top and
bottom of the domain.  A small perturbation is made to the center of
the current sheet at $t=0$ to begin reconnection.

\begin{figure}
\includegraphics[width=\textwidth]{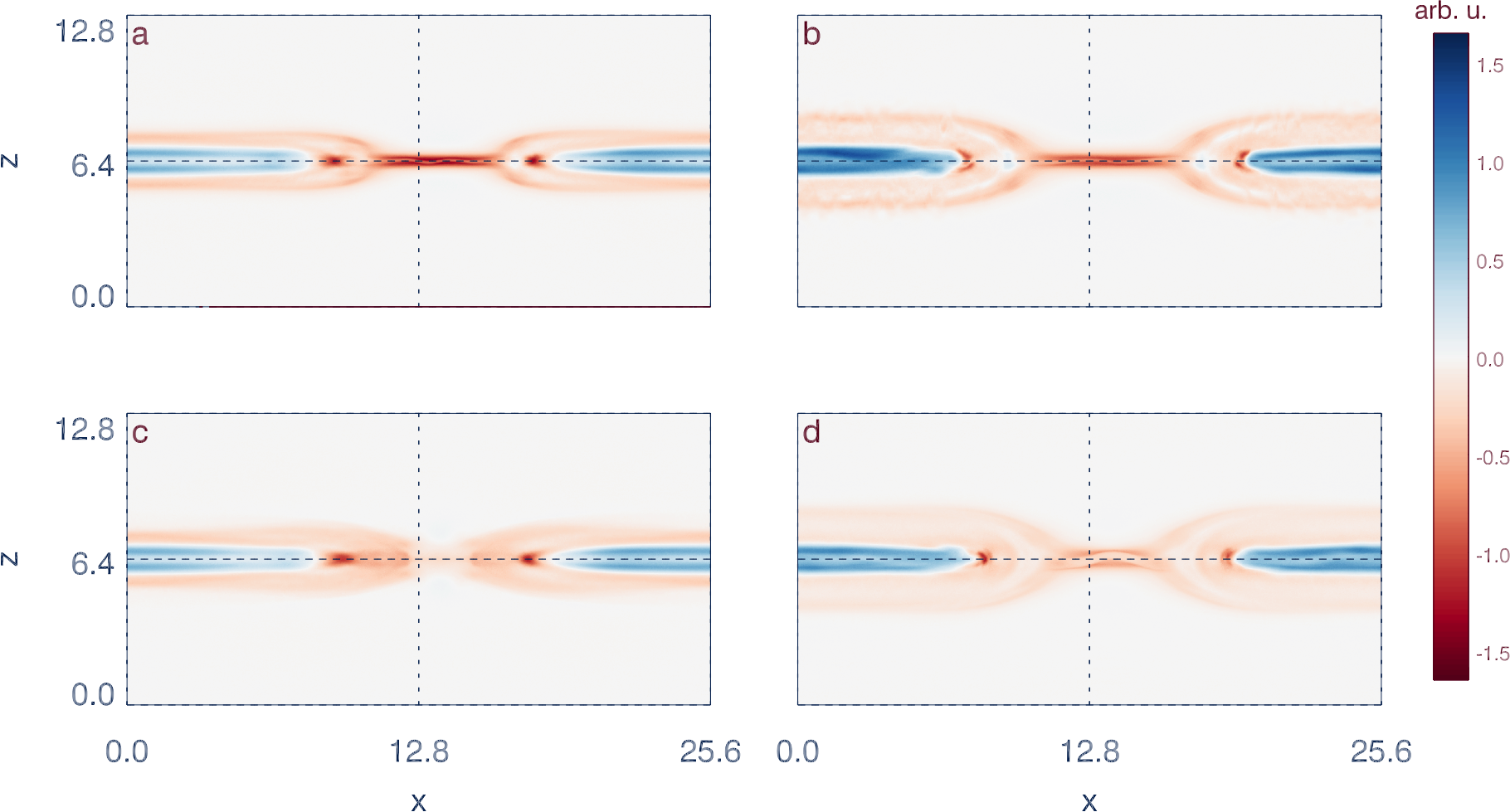}
\caption{\label{anti_gem} The out-of-plane electron current density
  $J_{ey}$ at four times for anti-parallel reconnection.  Panel (a)
  shows field lines and $t=20\Omega_{ci}^{-1}$.  Panel (b) shows
  $t=24\Omega_{ci}^{-1}$ when the forward evolution is stopped.  Panel
  (c) shows $t=20\Omega_{ci}^{-1}$ in the backwards evolution and
  panel (d) $t=12\Omega_{ci}^{-1}$.  [Associated dataset available at
    \protect\url{http://dx.doi.org/10.5281/zenodo.4608554}.]\cite{reversibility_fig5}}
\end{figure}

Figure \ref{anti_gem} displays the out-of-plane electron current density
$J_{ey}$ at four times during the run.  Panels a and b display the
system at $t=20\Omega_{ci}^{-1}$ and $t=24\Omega_{ci}^{-1}$ during the
forward time integration.  The expected features are present: slowly
inflowing plasma, growing islands of reconnected flux and Alfv\'enic
outflow jets (not shown).  At $t=24\Omega_{ci}^{-1}$, panel b, the run
is stopped and the backwards integration begun. Depending on the
details of the numerical implementation, small errors can be
introduced when starting the backwards integration.  However, tests
with multiple schemes taking more or less care to ensure exact
reversibility during the first few time steps yielded essentially
indistinguishable results.  In the following we simply set $\Delta t
\rightarrow -\Delta t$.  Panel c shows the system after it has been
integrated back to $t=20\Omega_{ci}^{-1}$ and should be compared to
panel a.  While some reversal has occurred -- in particular, the
island widths are similar in panels a and c -- the structure of the
central current layer is clearly different.  Panel d shows the system
after it has been evolved further backwards in time to
$t=12\Omega_{ci}^{-1}$.  If the system were fully reversible the
islands would continue to shrink as flux ``unreconnects'' and the
system would eventually contain a nearly uniform current layer.
Instead, reconnection has begun again, as seen in the increase in the
island size.  The overall morphology closely resembles the state shown
in panel b when the time reversal began.

\begin{figure}
\includegraphics[width=\textwidth]{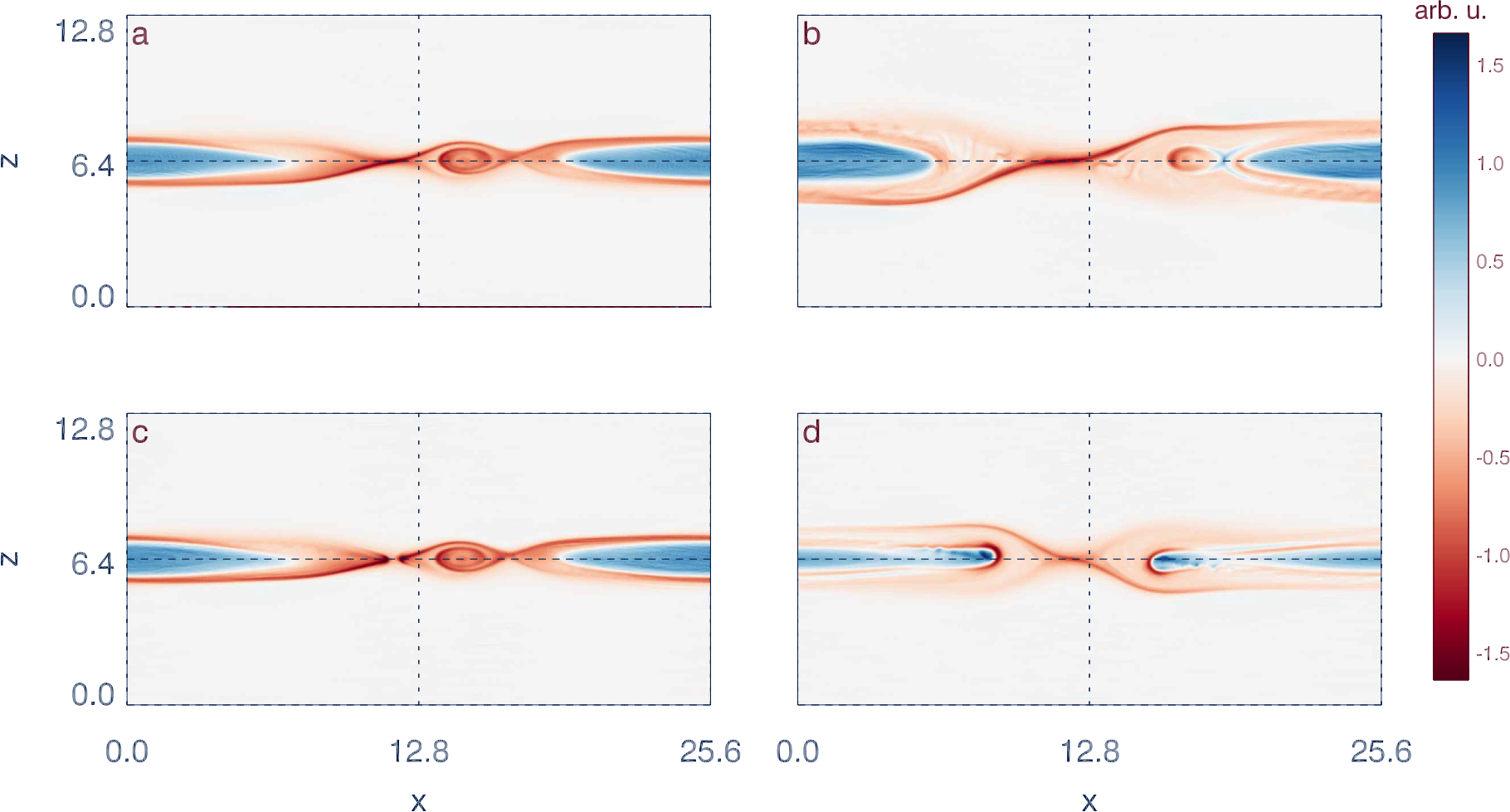}
\caption{\label{guide_gem} The out-of-plane electron current density
  $J_{ey}$ at four times for guide field reconnection with $B_y=1$ in
  the same format as Figure \ref{anti_gem}.  [Associated dataset
    available at
    \protect\url{http://dx.doi.org/10.5281/zenodo.4608560}.]\cite{reversibility_fig6}}
\end{figure}

Figure \ref{guide_gem} displays the out-of-plane current density at
four times for a system that includes an initial uniform guide field,
$B_y = 1$, but is otherwise identical to that shown in Figure
\ref{anti_gem}.  As before, panels a and b show $t=20\Omega_{ci}^{-1}$
and $t=24\Omega_{ci}^{-1}$ from the forward time integration.  In this
case a plasmoid forms in the the slightly narrower current layer near
the X-line and moves downstream, nearly fully reconnecting with the
flux in the island by the time of panel b.  As is typical in
guide-field reconnection, one of the separatrices, in this case the
one stretching from lower-left to upper-right, is stronger than the
other.  After the time shown in panel b the backwards integration
begins.  In panel c (which should be compared to panel a) the plasmoid
has emerged from the downstream island and propagated back towards the
X-line.  The two panels are nearly identical although small
differences can be seen, e.g., at $x\approx 11$ and $z\approx 6.4$.
However, as the reversed integration continues, reconnection
eventually begins again as can be seen in panel d.

A curious feature of the new phase of reconnection is that the strong
and weak separatrices switch (compare panels b and d).  The morphology
of the separatrices before the reversal is a well-understood byproduct
of guide-field reconnection and arises from the interaction of
$E_{\parallel}$ along the separatrices and streaming electrons
\cite{pritchett04a}.  As noted above, reversing the sign of time in
the equations governing the system's evolution (i.e., mapping
$t\rightarrow -t$) is mathematically equivalent to keeping $t$
unchanged and instead mapping $\mathbf{B}\rightarrow-\mathbf{B}$ and
$\mathbf{v}\rightarrow-\mathbf{v}$.  In other words, instead of
reversing the flow of time the system can be considered to reverse the
direction of the magnetic field and the velocity of every particle in
the system instantaneously.  Reversing the sign of $\mathbf{B}$ in
normal (non-time-reversed) guide-field reconnection reverses which
separatrices appear bright and so is consistent with the change in the
separatrices shown in panel d.

\begin{figure}
\includegraphics[width=0.8\textwidth]{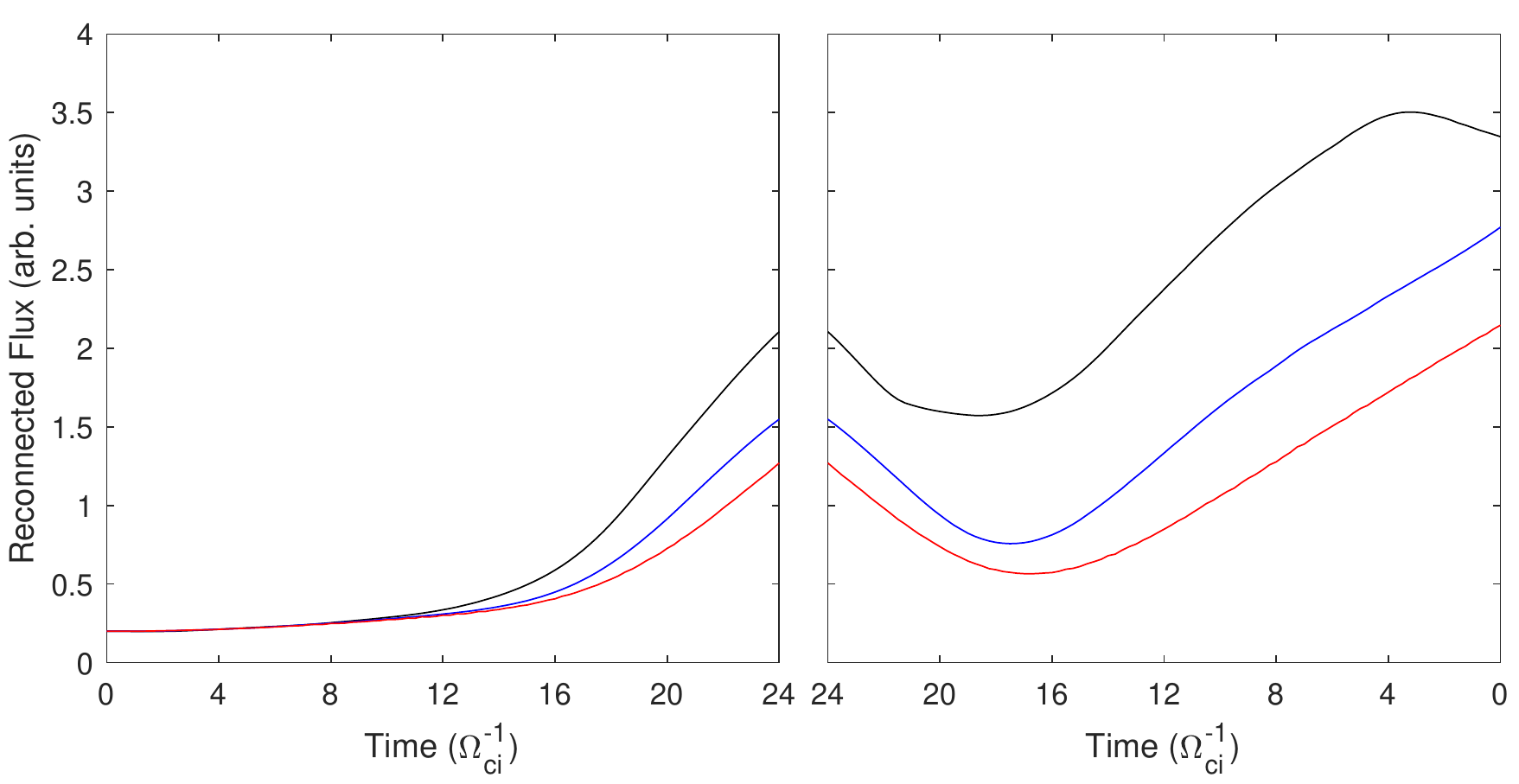}
\caption{\label{rates} Reconnected flux versus time for simulations
  with guide field 0 (black), 1 (blue) and 2 (red).  The left panel
  shows forward integration in time.  The right panel, the backward
  integration.  [Associated dataset available at
    \protect\url{http://dx.doi.org/10.5281/zenodo.4608562}.]\cite{reversibility_fig7}}
\end{figure}

Figure \ref{rates} shows the reconnected flux, as measured by the
difference in $A_y$ between the X-point and O-point, versus time for
three simulations: The two already discussed and an additional one
with $B_y=2$.  The slope of the curve gives the reconnection rate,
which is $\approx 0.1$ for all three simulations at the time of the
reversal.  The left panel shows the forward integration, with the
vertical offset at $t=0$ due to the small initial perturbation.  Each
simulation is run until $t=24$, the time shown in panel b of Figures
\ref{anti_gem} and \ref{guide_gem}.  The right panel displays the
result of the backward integration.  In each case the system retraces
it trajectory (``unreconnects'') immediately after the reversal, but
eventually stops and then begins to reconnect new flux.  As was the
case with the test particles discussed in Section \ref{trajs} the
systems with guide fields show better reversibility.

\begin{figure}
\includegraphics[width=0.8\textwidth]{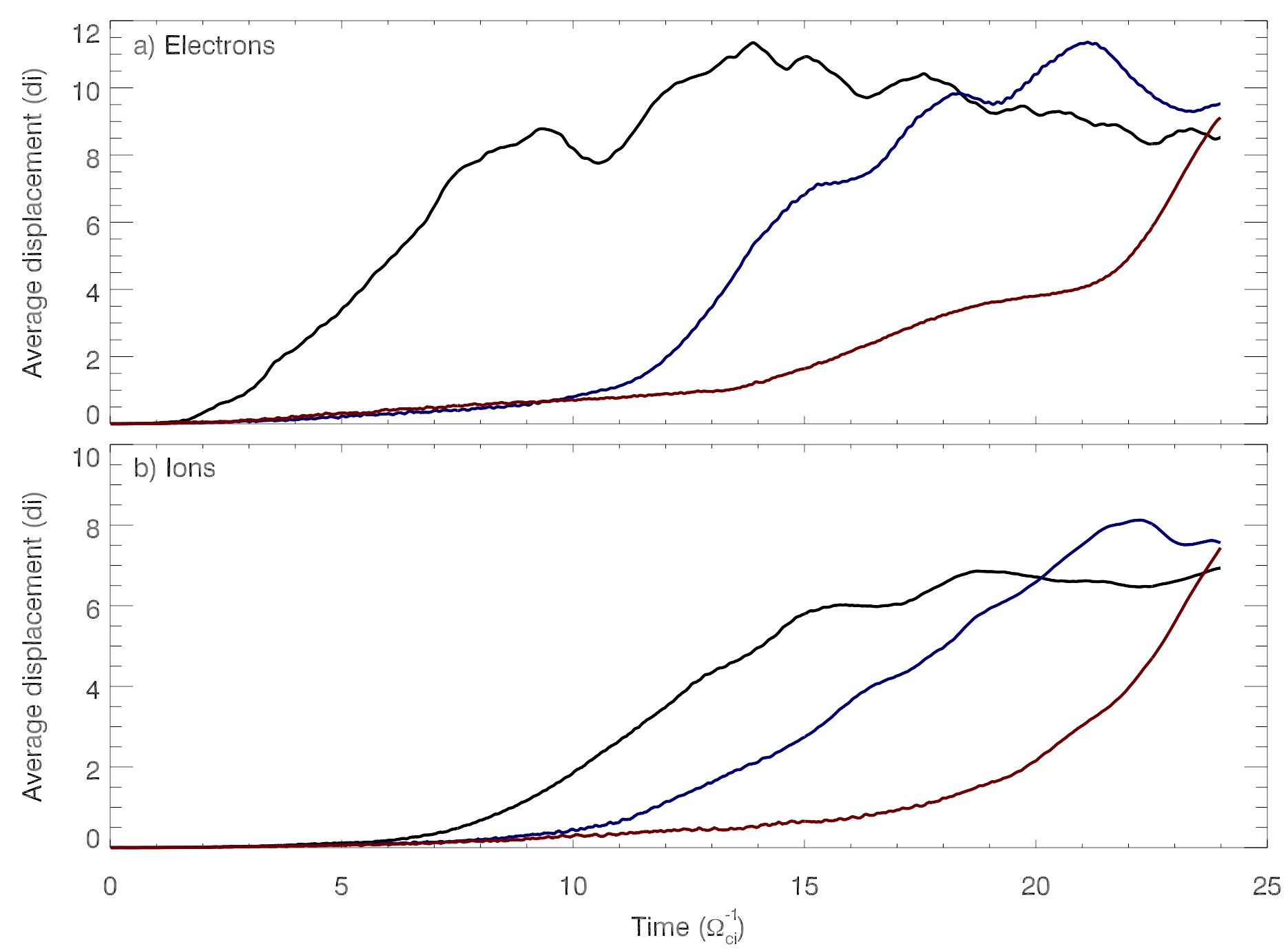}
\caption{\label{disp} Average displacement between forward and
  backward runs as a function of time for electrons (panel a) and ions
  (panel b).  The three curves correspond to runs with guide field 0
  (black), 1 (red), and 2 (blue).  [Associated dataset available at
    \protect\url{http://dx.doi.org/10.5281/zenodo.4608582}.]\cite{reversibility_fig8}}
\end{figure}

Unlike the test particles discussed in Section \ref{trajs}, particles
in PIC simulations move on trajectories that evolve based on
self-consistently determined electromagnetic fields.  In a perfectly
reversible simulation the trajectories from the forward and backward
integration would completely overlap but, as might be expected from
the irreversibility shown in Figure \ref{rates}, they do not here.
Figure \ref{disp} shows the average total displacement between the
forward and backward integration of fifty electrons (panel a) and ions
(panel b) as a function of time for reconnection with $B_g=0$ (black
curves), $B_g=1$ (red) and $B_g=2$ (blue).  The particles initially
occupied a small region just upstream of the reconnection current
sheet.  Most passed the separatrices during the simulation and were
located in the reconnection outflow when the backwards integration
began (this point in time denotes $t=0$ on the horizontal axis).
Because the system is periodic in the horizontal direction and the
particles cannot move too far upstream in the $y$ direction due to
canonical momentum conservation, the displacement between particles
cannot grow without bound.  Instead, in all cases the particles
initially track their dopplegangers closely before exponentially
diverging and then eventually reaching an asymptotic steady-state.  As
expected, the deviations occur substantially earlier when $B_g=0$ due
to passages through the demagnetized current sheet.  As the guide
field increases the correlation between the forwards and backwards
particles is maintained for longer periods, although eventually
divergence occurs for all three cases.  While the larger mass of ions
(panel b) means larger Larmor radii (and hence less magnetization than
electrons), it also increases the gyroperiod; the net result is that
the timescale for ion deviation is longer than that for electrons.

\section{Discussion}\label{disc}

Previous work \citep{ishizawa13a} has demonstrated fully reversible
magnetic reconnection in a gyrokinetic simulation of a collisionless
plasma.  The gyrokinetic equations effectively consider the case of
very strong guide field, i.e., a fully magnetized plasma, and hence
can be considered a limiting case of the work discussed here.  The
observed reversibility is consistent with the trend discussed above
that increasing the strength of the guide field increases the system's
reversibility.  The fluid nature of the gyrokinetic model caused the
noise levels in the simulations to be quite small (near machine
precision), although artificially adding noise led to poorer
reversibility.  Observations of reconnection have shown it to be
accompanied by small-scale electric-field fluctuations that have some
similarities to this type of noise \cite{ergun16a,ergun17a}.  It would
be of interest for future work to compare the characteristics
(amplitude, spectrum, etc.) of this noise to that in different
simulation models in order to assess its possible impact on reversibility.

Irreversibility is closely tied to the question of dissipation of
energy.  Circumstantial evidence for dissipation during reconnection
comes in the form of the observed increase of the downstream plasma
temperature \cite{phan13b,haggerty15a}.  However, definitively proving
the existence of irreversible dissipation is not straightforward.  A
frequently used measure, the identification of regions where
$\mathbf{J}\boldsymbol{\cdot}\mathbf{E} > 0$ (where $\mathbf{E}$ is
measured in the frame of the electrons \cite{zenitani11a}), can be
misleading since reversible processes can generate such signals.  On
the other hand, such signals can be tied to the the development of
complex structures in phase space. As the complexity increases, weaker
and weaker non-ideal processes are sufficient to trigger irreversible
heating \cite{swisdak18a}.

Determining whether a process is reversible is equivalent to asking
whether the entropy remains constant.  The generation of entropy in
PIC simulations has been a subject of recent study \cite{liang19a}.
The kinetic entropy was calculated for collisionless 2.5D
anti-parallel reconnection in a closed system.  For a simulation with
excellent conservation of the total energy, the authors found a
monotonic increase in the total entropy of the system, albeit at a low
level ($\approx 3\%$ for the entire run).  Since the system lacked
physical collisions, the entropy increase was attributed to numerical
effects.  Later work \cite{liang20a} modelled the entropy increase as
an effective numerical collisionality, with the collision frequency
dependent on such numerical factors as the resolution, timestep, and
number of particles.  When considered in the context of this work,
these results suggest that reversibility in magnetic reconnection
behaves in a manner analogous to Landau damping.  In principle every
particle in the system follows a trajectory given by the
collisionless, reversible equations of motion while the system itself
follows a narrow trajectory through a high-dimensional phase space.
However, during reconnection the orbits of particles passing near
magnetic separatrices and magnetic nulls are sensitive to small
perturbations.  Deviations from collisionless trajectories (e.g.,
arising from numerical errors in a simulation or a non-zero effective
collisionality in a real system) pushes the plasma off its narrow path
in phase space onto an irreversible state.  The size of the
perturbation needed to effect this change varies, although we have
shown that a larger guide field is linked to higher reversibility (and
hence smaller entropy increase and energy dissipations).

Several open questions remain. To what degree does reversibility
depend on other plasma parameters (e.g., three-dimensionality,
$\beta$, the magnetization paramater $\sigma$ that parameterizes
relativistic reconnection \cite{guo14a}, or asymmetry across the
reconnecting current sheet)?  Is it possible for PIC simulations to
run in the truly collisionless regime for general reconnection
configurations and hence exhbit true reversible behavior?  If not,
does this have implications for the widely accepted notion that PIC
simulations provide a nearly complete picture of magnetic
reconnection?  And finally, is there a definitive measure of
irreversibility and entropy generation that can be evaluated with
data obtainable by either spacecraft or laboratory experiments?

\begin{acknowledgments}
The authors acknowledge support from NSF grant No. PHY1805829 and NASA
grant 80NSSC19K0396.  This research used resources of the National
Energy Research Scientific Computing Center, a DOE Office of Science
User Facility supported by the Office of Science of the
U.S. Department of Energy under Contract No. DE-AC02-
05CH11231. Simulation data are available upon request.
\end{acknowledgments}

%\bibliography{paper,datasets}
%\bibliographystyle{elsarticle-num}

\end{document}